\documentclass[copyright,creativecommons]{eptcs}
\providecommand{\event}{WWV 2011} % Name of the event you are submitting to

\title{Automated Functional Testing based on the Navigation of Web Applications}
\author{Boni Garc\'ia \qquad\qquad Juan C. Due\~nas
\institute{Departamento de Ingenier\'ia de Sistemas Telem\'aticos \\ ETSI Telecomunicaci\'on - Universidad Polit\'ecnica de Madrid \\Avda. Complutense 30, 28040 \\ Madrid, Spain}
\email{bgarcia@dit.upm.es \qquad\qquad jcduenas@dit.upm.es}
}
\def\titlerunning{Automated Functional Testing based on the Navigation of Web Applications}
\def\authorrunning{B. Garc\'ia \& Juan C. Due\~nas}

\usepackage{graphicx}
\usepackage{moreverb}
\usepackage{pstricks-add}
\usepackage{listings}
\usepackage{color}
\usepackage{textcomp}
\usepackage{xcolor}
\usepackage{courier}
\usepackage{amssymb}

\usepackage{lstcustom}
\usepackage{pseudocode}

\definecolor{darkgreen}{named}{green}
\definecolor{darkblue}{named}{blue}
\definecolor{darkred}{named}{red}
\definecolor{grau}{named}{gray}
\definecolor{listinggray}{gray}{0.9}
\definecolor{lbcolor}{rgb}{0.9,0.9,0.9}
\begin{document}
\maketitle

\begin{abstract}
Web applications are becoming more and more complex. Testing such applications is an intricate hard and time-consuming activity. Therefore, testing is often poorly performed or skipped by practitioners. Test automation can help to avoid this situation. Hence, this paper presents a novel approach to perform automated software testing for web applications based on its navigation. On the one hand, web navigation is the process of traversing a web application using a browser. On the other hand, functional requirements are actions that an application must do. Therefore, the evaluation of the correct navigation of web applications results in the assessment of the specified functional requirements. The proposed method to perform the automation is done in four levels: test case generation, test data derivation, test case execution, and test case reporting.  This method is driven by three kinds of inputs: i) UML models; ii) Selenium scripts; iii) XML files. We have implemented our approach in an open-source testing framework named Automatic Testing Platform. The validation of this work has been carried out by means of a case study, in which the target is a real invoice management system developed using a model-driven approach.\end{abstract}

\section{Introduction}
\label{s1}

The World Wide Web (or simple the Web) has become one of the most influential instruments not only in computing but in the history of mankind \cite{1}. The development of web applications has been in general ad hoc, resulting in poor-quality applications. As the reliance on larger and more complex web applications increases so does the need for using methodologies and guidelines to develop applications that are delivered on time, within budget, and with a high level of quality. 

Software testing is the main technique to ensure quality and finding bugs. It is in general a difficult and time-consuming task. Web testing may be even more difficult, due to the peculiarities of such applications. A significant conclusion has been reached in the survey of web testing depicted in  \cite{2}: \textit{``further research efforts should be spent to define and assess the effectiveness of testing models, methods, techniques and tools that combine traditional testing approaches with new and specific ones''}. Following this statement, this piece of research presents an approach to test web applications by automating its navigation. This approach allows saving time on testing effort, since it is automated in different levels. Firstly, unit test cases are generated for each path in the navigation of the web site. Secondly, depending on the data handled by the System Under Test (SUT), test data is generated and stored in a spread-sheet. These data is used later by the created unit test cases. Thirdly, test case execution is performed using Selenium, which allows running the web interaction using a real browser. Finally, a complete test report is generated of each unit test case, which corresponds to each path in the SUT navigation.

The remainder of this paper is structured as follows. Section~\ref{s2} introduces the context in which this research has been performed (i.e. automated software testing, web testing and modeling, and graph theory). Section~\ref{s4} shows a comparative study of the existing choices to find the paths in the web navigation. This section also presents a laboratory experiment carried out in order to select the algorithm to be uses in the proposed approach. Section~\ref{s3} presents the proposed methodology to perform automated web testing. Section~\ref{s5} describes how the proposed method has been implemented in an open-source tool named Automatic Testing Platform (ATP). Section~\ref{s6} details the validation of the presented approach by means of an industrial case study, in which ATP has been employed to test an invoice management web system. Finally section~\ref{s7} presents the reached conclusions.

\section{Background}
\label{s2}

\subsection{Automated Software Testing}
\textbf{Software testing} is the main activity performed for evaluating product quality, and for improving it, by identifying defects in software-intensive systems \cite{3}. Manual testing is usually a hard and time-consuming and expensive activity. Some studies shows that testing is considered as one of the most costly development processes, sometimes exceeding fifty per cent of total development costs \cite{4}. Consequently, testing is often poorly performed or skipped by practitioners. This situation suggests industry-wide deficiency in testing, and automated testing is proposed as one possible solution to overcome this problem \cite{5}. A definition of Automated Software Testing (AST) can be found in \cite{7} as the \textit{``Application and implementation of software technology throughout the entire Software Testing Lifecycle (STL) with the goal to improve efficiencies and effectiveness''}.

AST is most effective when implemented within a framework. Testing frameworks may be defined as a set of abstract concepts, processes, procedures and environment in which automated tests will be designed, created and implemented. According to the Automated Testing Institute  (ATI)\footnote{\href{http://www.automatedtestinginstitute.com/}{http://www.automatedtestinginstitute.com/}}, there are three different generations of AST frameworks. The \textbf{1\textsuperscript{st} generation} is comprised of the linear approach to automated testing. This approach is typically driven by the use of the Record \& Playback (R\&P) method. It is carried out firstly recording the linear scripts corresponding with actions performed in the application (record). After that, the automation stage can be done repeating the record while exercising the SUT (playback). The \textbf{2\textsuperscript{nd} generation} comprises two kinds of frameworks: the data-driven and functional decomposition. Frameworks built on data-driven scripting use test data typically stored in a database of file external. Functional decomposition refers to the process of producing modular components in such a way that automated test scripts can be constructed to achieve a testing objective by combining these existing components. The \textbf{3\textsuperscript{rd} generation} includes the keyword-driven and model-based frameworks. The keyword-driven frameworks process automated tests that are developed with a vocabulary of keywords. These keywords are associated with functions that are interpreted with application-specific data. The automated scripts execute the interpreted statements in the SUT. The model-based frameworks go beyond creating automated tests in a semi-intelligent manner. Model-Based Testing (MBT) is the software testing technique where test cases are derived in whole or in part from a model that describes some aspects of the SUT \cite{9}. 

\subsection{Web Testing and Modeling}
\label{s22}
Web applications follow a client-server application protocol. The web client (using a web browser, such as Explorer, Opera, Safari, Firefox or Chrome) sends an HTTP request through a TCP-IP network (typically the Internet) to a web server. The server receives this request and determines the page, which usually contains some script language to connect which a database server. A middleware component connects the web server with the database to inform about query and get the requested data. This data is used to generate an HTML page, which is sent back to the client in form of a HTTP response. 

\textbf{Web testing} consists of executing the application using combinations of input and state to reveal failures. These failures are caused by faults in the running environment or in the web application itself. The running environment mainly affects the non-functional requirements of a web application, while the web application is responsible for the functional requirements. Web applications are difficult to test, due to their peculiarities \cite{2}: i) A wide number of users distributed all over the world accessing concurrently. ii) Heterogeneous execution environments (different hardware, network connections, operating systems, web servers and browsers). iii) Heterogeneous nature because of different technologies, programming languages, and components. iv) Dynamic nature, since web pages can be generated at run time according to user inputs and server status.

Regarding \textbf{web modeling}, in some cases new models have been proposed while in other cases existing modeling techniques have been adapted from other software domains \cite{13}. The de-facto notation standard for modeling is UML (Unified Modeling Language). The standard diagrams in UML 2.0 are the following \cite{14}: uses cases, activity, classes, sequence, interaction, communication, object, state machine, composite, deployment, package, and timing. Hence, UML 2.0 does not provide in a standard way any diagrams to model some specific aspects of the web applications. For that reason, specific UML extensions have been created, for example the

\begin{itemize}
\item UML-based Web Engineering (UWE)\footnote{\href{http://uwe.pst.ifi.lmu.de/}{http://uwe.pst.ifi.lmu.de/}} is a software engineering approach aiming to cover the whole life-cycle of web application development \cite{15}. It is based on the Unified Process (UP) \cite{16}, and defines a UML notation by means of an UML profile.
\item W2000 is an approach that also extends UML notation to model multimedia elements. These multimedia elements are inherited from HDM (Hypermedia Design Model) \cite{17}.
\item Web Modeling Language (WebML)\footnote{\href{http://www.webml.org/}{http://www.webml.org/}} is a high-level specification language for designing complex web applications. It offers a visual both in Entity-Relationship and UML, although UML is preferred by the authors \cite{18}.
\item Navigational Development Techniques (NDT)\footnote{\href{http://iwt2.org/}{http://iwt2.org/}} is a methodological approach oriented to the web engineering. It is mainly focused on the requirements and the analysis phases using the model-driven paradigm \cite{19}.
\end{itemize}

\subsection{Graph Theory}
In mathematics and computer science, graph theory is the study of graphs. A graph is the abstract representation of a set of vertices (vertex or nodes) connected by arcs (edges or links). A graph is a pair $G=(V,E)$ of sets such that the elements of $V$ are vertex and the elements of $E$ are the edges. The usual way to picture a graph is by drawing a dot for each vertex and joining two of these dots by a line if the corresponding two vertices form an edge \cite{22}. 

On one hand, a graph in with the edges have no orientation is known as undirected graph. On the other hand, if the edges have orientation, the graph is known as \textbf{directed graph (digraph)} \cite{23}. A digraph is acyclic if it has no cycle. A digraph is strongly connected (or, just, strong) if every vertex is reachable from every other vertex, i.e. there is a path from each vertex in the graph to every other vertex. A multigraph is a graph in which is permitted having multiple edges (two or more edges that are incident to the same two vertices) and/or loops (edge that connects a vertex to itself). If the multigraph is directed, then is known as \textbf{multidigraph}. In a weighted graph a number is assigned to each edge. This number (weight) could represent costs, lengths and so on.

A \textbf{path} is a graph such that from each of its vertex there is an edge to the next vertex in the sequence. If the start node is the same than the end node, then the path is known as cycle. A walk is a path in which nodes or links may be repeated. A circuit is closed walk.

\section{Finding the Paths in a Multidigraph}
\label{s4}

We will use graph theory to represent and work with the navigation of web applications. Therefore, a web site can be modeled by means of a finite multidigraph, that is, a finite directed graph (finite set of web pages and nodes) in which multiple edges and/or loops are allowed. Given a multidigraph, we need a method or algorithm to find its independent paths. The coverage criteria in this path decomposition is that each edge is \textbf{traversed at least once}. This condition also implies that each vertex is visited at least once too. This section studies the algorithms and methods found in the literature to solve this problem. Some discussion is provided in order to select the best option.

\textbf{Graph traversal} is the facility to move through a structure visiting every vertex once. There are two possible traversal methods for a graph: Breadth-First Search (BFS) and Depth-First Search (DFS) \cite{25}. BFS visits all the vertex, beginning with a specified start. No vertex is visited more than once. BFS makes use of a queue (First-In First-Out, FIFO) data structure. DFS works in a similar way, except that the neighbors of each visited vertex are added to a stack (Last-In, First-Out, LIFO) data structure. \textbf{Traveling Salesman Problem (TSP)} tries to find the most efficient (i.e., least total distance) cycle through each of each vertex of a graph \cite{24}. TSP is a variation of the Hamiltonian tour problem (to find a cycle that visits each vertex exactly once in a graph), and it belongs to the class of NP-hard problems. \textbf{The Shortest Path Problem (SPP)} is the problem of finding a path between two nodes within a graph such that the sum of the weights of its constituent edges is minimized \cite{25}. The main algorithms employed in the different categories of SPP are: Dijkstra, Bellman-Ford, A* (pronounced ``A star''), and Floyd-Warshall.

\textbf{The Chinese Postman Problem (CPP)}, also known as the postman tour or route inspection problem, is the problem of finding a shortest circuit that visits every edge of a graph at least once, i.e. the Chinese Postman Tour (CPT). Finding an optimal solution of these problems is NP-complete \cite{26}. Thimbleby proposes a solution for CPP in form of deterministic algorithm in \cite{27}, providing an executable Java to solve this problem. The constraint imposed by this algorithm is that the input digraph has to be strongly connected with no negative weight cycles. It considers a graph as a collection of arcs $<label,i,j,c>$, where label is an identifier for an arc from vertex $i$ to $j$, and $c$ the cost associated with it. 

\textbf{The node reduction algorithm} \cite{4} finds out the path between two nodes, typically the entry and exit nodes by reducing the rest of graph connecting these nodes. It employs graph algebra to achieve this goal. The multiplicative operator in graph algebra means concatenation: if edge a is followed by edge b, their product is $a \cdot b$ (path product). The additive operator is selection: if either edge a or edge b can be taken, their sum is $a+b$. A path expression contains path products and zero or more additive operators, and are usually represented by upper case letters (e.g. $A = a \cdot b$) \cite{28}. Finally, in graph algebra it is usually employed the graph matrix representation, which is a square array with one row and one column for every node in the graph. Each row-column combination corresponds to a relation between the node corresponding to the row and the node corresponding to the column \cite{4}. The node reduction algorithm has basically two steps: i) remove self-loops (any node n that has an edge to itself); ii) eliminate intermediate nodes and replacing it with a set of equivalent links. An example of this algorithm is detailed below, and it has been employed for web navigation in \cite{29}.

None of the methods before fits exactly in the problem at hand: to select the different path within a graph. BFS and DFS algorithms traverse each vertex within a graph, but they do not ensure that each edge is visited at least once. This applied to web navigation is not acceptable, due to the fact that we need to visit each web link. The same issue happens with TSP: Hamiltonian tours have nothing to with edges but vertices coverage. The different algorithms of SPP are not useful in this domain due to the fact that it looks for the shortest path between nodes. CPP fits exactly with the objective of 100\% edge coverage, but it has a strong constraint that cannot be ensured for any multidigraph modeling web navigations: they should be strong connected. Node reduction could be an alternative, but we cannot suppose that web navigation has always an exit page.

\begin{figure}
  \caption{Digraph Example (Original and Strongly Connected)}
  \label{f1}
  \begin{center}
  	\includegraphics[width=260px]{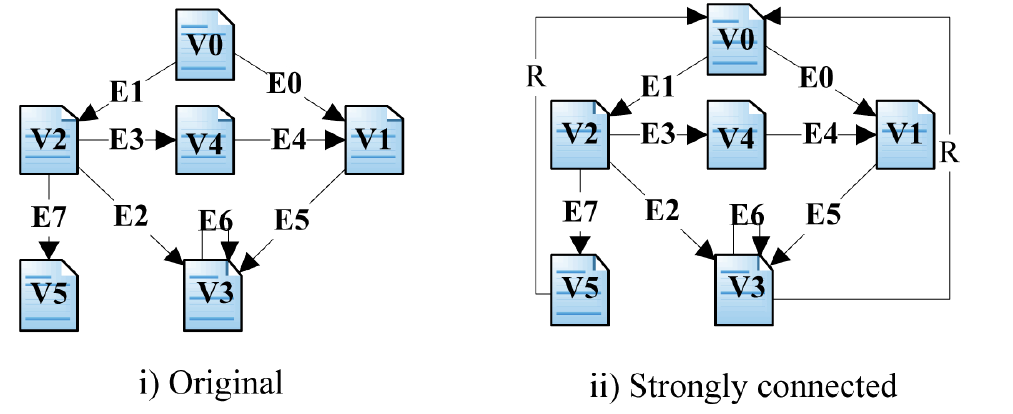}
  \end{center}
\end{figure}

Nevertheless, CPP and node reduction can be modified to solve the problem. Consider the digraph labeled as ``i) Original'' in Figure \ref{f1}. A simple and effective way to convert this graph in strongly connected is by adding virtual links from the leaf nodes (those with no out links), connecting them with the start node (``home'' in the navigation). These virtual links are labeled with ``R'', which means ``reset''. These links will be substituted by additive operator when reducing the graph to its paths. The new equivalent digraph is shown in Figure \ref{f1}, labeled as ``ii) Strongly connected''.

\begin{figure}
  \caption{Node Reduction Solution for the Proposed Example}
  \label{f2}
  \begin{center}
  	\includegraphics[width=300px]{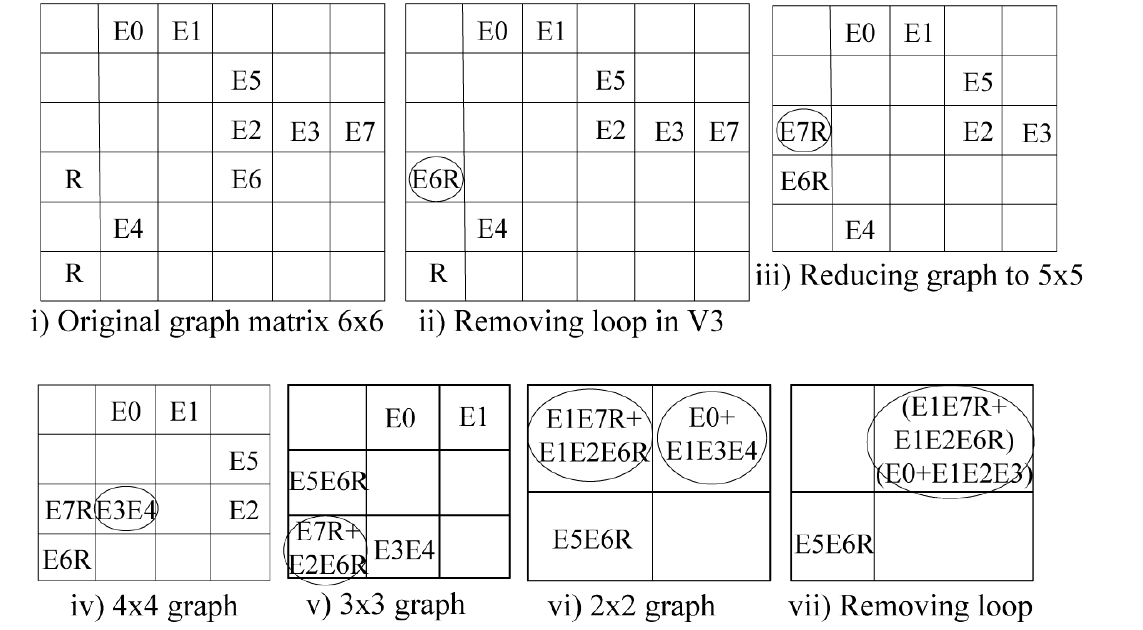}
  \end{center}
\end{figure}

In this new situation, CPP suits the problem of checking web sites: a vertex is a page, an arc is a link, the label of the arc could be hot text or a URL, and the weight represents a cost (e.g. estimated in seconds of the user checking the link). The goal is to determine a list of labels that, in order, constitute an optimal CPT, i.e. the shortest tour with few repeated street visits to cover every edge in the graph. The cost of a CPT is defined as the total arc weight, summed along the circuit. The optimal test sequence for this web site is therefore an open CPT. Assigning a weight of 1 for each link and applying the CPP algorithm given in \cite{27} to the proposed example, the resulting path expression is the following:

\begin{lstlisting}[mathescape]
$E1 \cdot E3 \cdot E4 \cdot E5 \cdot R \cdot E1 \cdot E7 \cdot R \cdot E1 \cdot E2 \cdot E6 \cdot R \cdot E0 \cdot R = E1 \cdot E3 \cdot E4 \cdot E5 + E1 \cdot E7 + E1 \cdot E2 \cdot E6 + E0$
\end{lstlisting}

That is, four different paths with a total cost of 10 links. Moreover, node reduction can be applied to the strongly connected graph in order to reduce the equivalence graph matrix. The complete explanation of how this process is done can be found in \cite{4}, and the resolution for this example it is illustrated in Figure \ref{f2}. Therefore, the resulting path expression of this application of the node reduction algorithm to the proposed example is:

\begin{lstlisting}[mathescape]
$(E1 \cdot E7 \cdot R + E1\cdot E2 \cdot E6 \cdot R) \cdot (E0 + E1 \cdot E2 \cdot E3) \cdot E5 \cdot E6 \cdot R = E1 \cdot E7 + E0 \cdot E5 \cdot E6 + E1 \cdot E3 \cdot E4 \cdot E5 \cdot E5 + E1 \cdot E2 \cdot E6$
\end{lstlisting}

\begin{figure}
  \caption{Node Reduction vs. CPP Costs}
  \label{f3}
  \begin{center}
  	\includegraphics[width=270px]{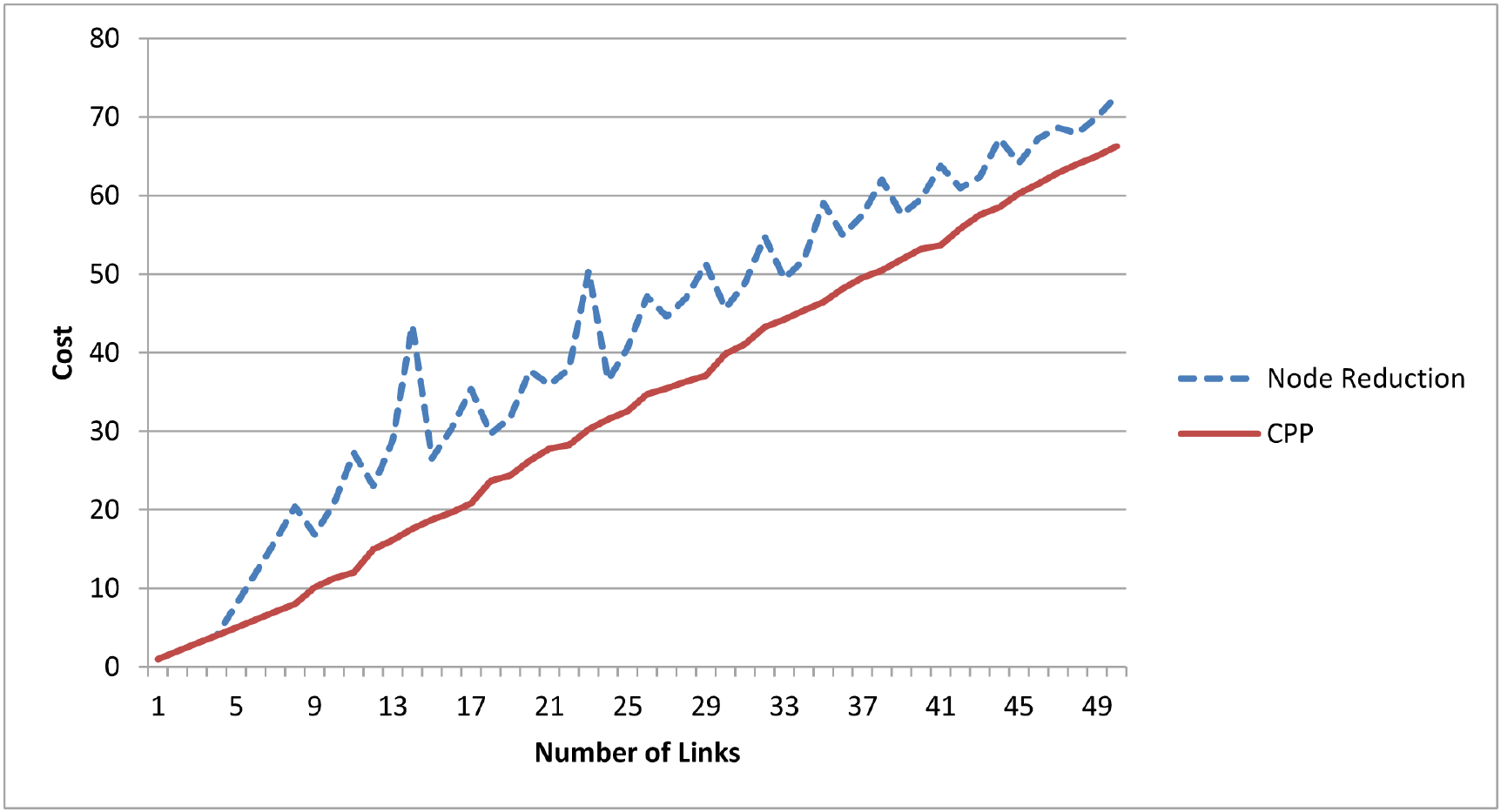}
  \end{center}
\end{figure}

\begin{figure}
  \caption{Node Reduction vs. CPP Time}
  \label{f4}
  \begin{center}
  	\includegraphics[width=290px]{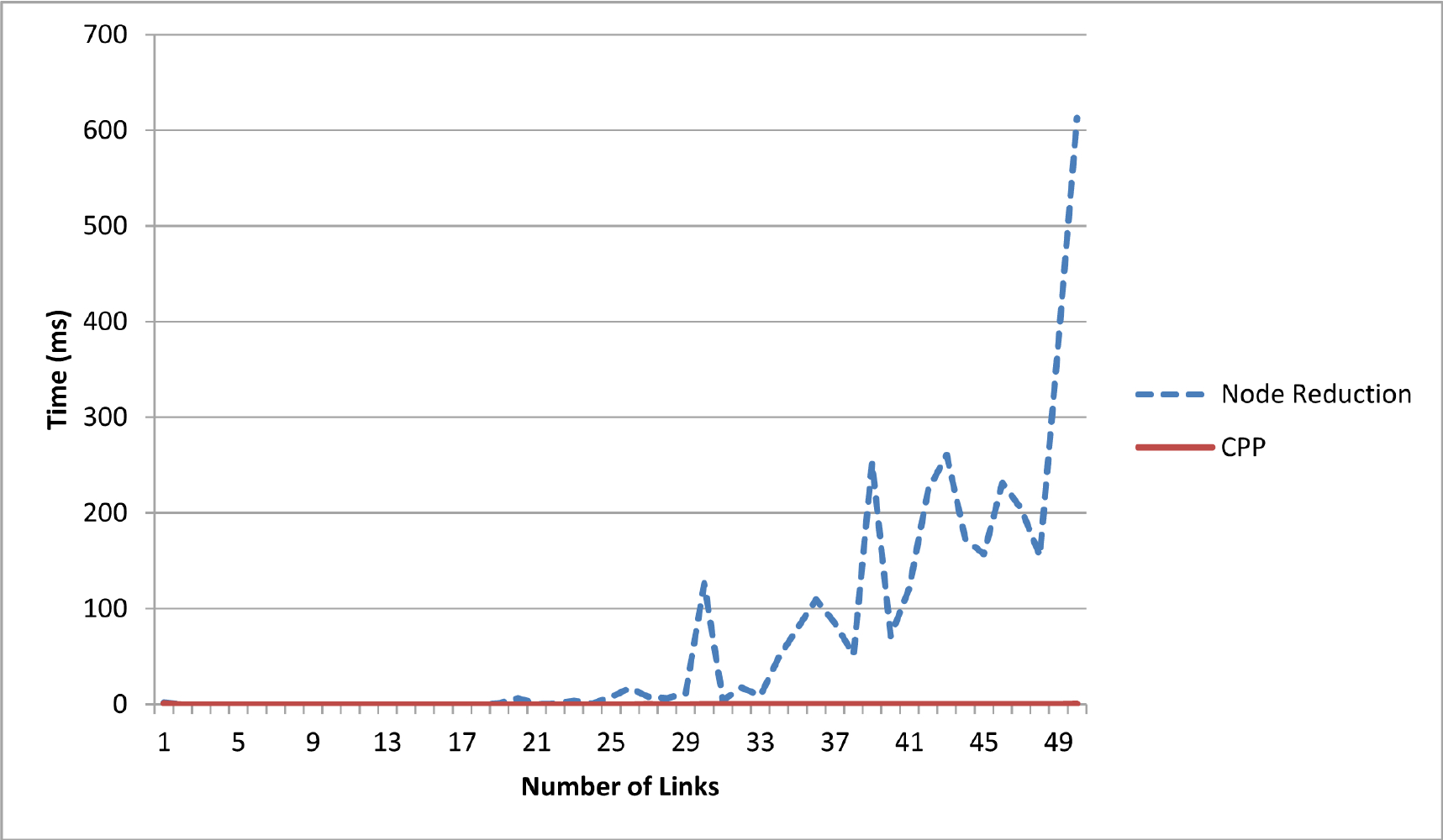}
  \end{center}
\end{figure}

That is, four different paths with a total cost of 13 links. It is a quite similar solution than the one provided by CPP (10 vs. 13 links). This fact suggests that CPP gives better results than node reduction. In order to ensure this statement, we have made a laboratory experiment. The experiment will consist on the comparison between node reduction and CPP, using random multidigraphs (i.e., with loops and multiple edges). These graphs have been created using an incremental number of links (from 1 to 50). For each digraph node reduction and CPP will be executed, comparing its cost (number of links employed in the resulting set of paths), and also the computation time (milliseconds in achieve the solution). This experiment has been carried out in a PC Intel Core2 Quad (2.66 GHz) with 4 GB of RAM memory. It has been repeated 100 times, and the mean of the values (cost and time) is shown in Figure \ref{f3} and \ref{f4}. 


CPP has a better behavior than node reduction because it is more linear. It always has a better cost solution than node reduction (Figure \ref{f3}). In addition, the resolution time is higher and higher in node reduction while CPP ends always in a few of milliseconds (Figure \ref{f4}). All in all, CPP is the selected algorithm to find out the set of path in a digraph in our method.

\section{Proposal Statement}
\label{s3}

Di Lucca and Fasolino draw an important conclusion about functional testing for web applications \cite{2}: \textit{``As to the functional testing, existing tools main contribution is limited to manage test case suites manually created, and to match the test case results with respect to a manually created oracle. Therefore, greater support to automatic test case generation would be needed to enhance the practice of testing Web applications''}. Following this statement, we propose an approach to perform AST for web applications based on its navigation. 

To achieve automated functional testing, requirements should be described in a form that can be understood by software programs. Hence, the first way we propose to model web navigation in order to automate the testing process is by means of \textbf{UML 2.0 diagrams}, concretely the following:

\begin{itemize}
\item Use case diagram. These diagrams offer a perspective of the functional requirements of the application interaction with the actors.
\item Activity diagrams. These diagrams describe the flow within a use case. Therefore, each activity diagram will describe the navigation structure of the SUT.
\item Presentation diagrams. These diagrams models the data which is handle by the web application. Due to the fact that UML 2.0. do no implement this feature, it is required a UML profile which enhances the syntax of standard UML 2.0 to achieve modeling of web pages. 
%Some alternatives have been presented in section \ref{s22}. Attending to its comparative shown in Table \ref{t1}, the best option is NDT, so it will be employed in our proposal.
\end{itemize}

As depicted in Section \ref{s22}, presentation diagrams are not standard in UML 2.0. Therefore, we need to use one of the specific UML extensions for web applications depicted in that section. Recent research shows that software project success is directly tied to requirement quality \cite{20}. Requirement Engineering (RE) involves all lifecycle activities devoted to identification and analysis of user requirements, documentation of the requirements as specification, and validation of the documented requirements against user needs. Thus, we are going to compare the presented UML-based technologies using which types of requirements are handled by each approach \cite{21}. Table \ref{t1} presents this comparison. Each column shows whether or not the technology manages the following type of requirements: i) data requirements (also known as conceptual requirements, establishes how information is stored and administrated); ii) user interface (interaction requirements); iii) navigation (users' navigation needs); iv) personalization (customization, describing how requirements are dynamically adaptable); v) transactional; vi) non-functional. Having seen these results, NDT seems to be the choice to model web applications since it covers each kind of the requirements studied. Therefore, the UML models to guide the MBT approach of this paper will be based on NDT.

\begin{table}
  \caption{UML-Based Web Modelling Technologies}
  \label{t1}
  \begin{center}
    \begin{tabular}{ | l | c | c | c | c | c | c | } \hline 
\textbf{} & \textbf{Data}  & \textbf{UI}  & \textbf{Navigation}  & \textbf{Personalization}  & \textbf{Transactional}  & \textbf{Non-functional} \\ \hline
UWE & $\checkmark$ & $\checkmark$ & $\checkmark$ &  &  &   \\ \hline
W2000 & $\checkmark$ & $\checkmark$ & $\checkmark$ & $\checkmark$ &  & $\checkmark$  \\ \hline
WebML &  &  &  & $\checkmark$ & $\checkmark$ &   \\ \hline
NDT & $\checkmark$ & $\checkmark$ & $\checkmark$& $\checkmark$ & $\checkmark$ & $\checkmark$\\ \hline
    \end{tabular}
  \end{center}
\end{table}

States in activity diagrams are connected by links. These links are characterized by a label called guard. This guard describes how web state changes and it will be used to describe the involved HTML elements in the transitions between states. A transition can be composed by several atomic actions. In order to be able to describe this behavior, the guard of the activity diagrams will follow the notation depicted below:

\begin{lstlisting}[mathescape]
[$target_{1}$,$event_{1}$,<$key_{1}$> ; $target_{2}$,$event_{2}$,<$key_{2}$> ; ... $target_{n}$,$event_{n}$,<$key_{n}$>] 
\end{lstlisting}

The meaning of these fields is the following:

\begin{itemize}
\item Target: Identifier of the HTML target element. In order to translate the target element, the following procedure will be used:

%\begin{lstlisting}[language=Ant,caption={Algorithm to find locators},label=l1]
\begin{lstlisting}[language=Ant]
Function LocateHTMLElement(Target)
	Found = nothing
	For each frame in the frameset (if frames exist)
		For each HTML element in the frame
			Found = Look for Target in the id/name/value attribute
			If Not Found
				Found = Look for Target as text
				If Not Found
					Found = Execute Target as XPath expression
				End If					
			End If
		End For
	End For
	Return Found
End Function
\end{lstlisting}

\item Event: Literal that describes the action performed. These literals are based on the DOM event specified by the W3C\footnote{\href{http://www.w3.org/TR/DOM-Level-2-Events/events.html}{http://www.w3.org/TR/DOM-Level-2-Events/events.html}}, i.e. \textit{click}, \textit{dblclik}, \textit{keypress}, \textit{keydown}, \textit{keyup}, \textit{mousedown}, \textit{mousemove}, \textit{mouseout}, \textit{mouseover}, and \textit{mouseup}.
\item Key: Optional field containing the button that triggers the key events.
\end{itemize}

The second way of modeling the web navigation will be using the \textbf{R\&P approach}, which is a useful way to represent the structure of a web application by recordings interactions with the application trough the browser. This method is more agile than UML, since the application can be developed avoiding the formal design phase. 

Halfway between the UML models and R\&P, we have created a syntax-neutral way of modelling the navigation using a specific created \textbf{XML notation}. XML (Extensible Markup Language) provides an easy way to store and share information. To provide the formal declaration of this XML format, XML Schema language (also known as XML Schema Definition, XSD) will be employed to perform the formalization of the navigation constraints.

\begin{figure}
  \caption{XSD Graphic Representation}
  \label{f5}
  \begin{center}
  	\includegraphics[width=260px]{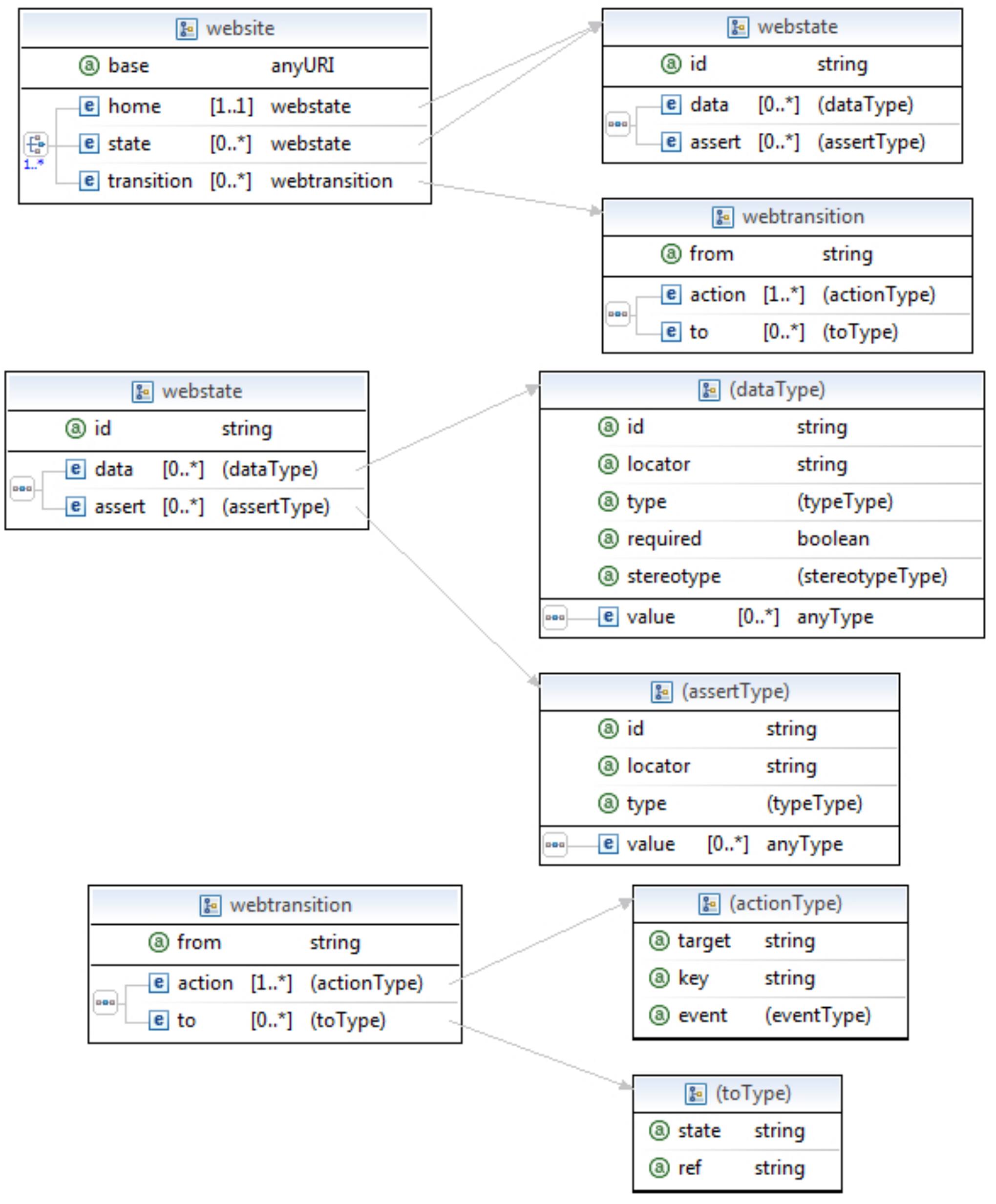}
  \end{center}
\end{figure}

This XSD schema defines a website as a collection of states (pages) and transitions (links). The initial page is called home, and it is unique. In addition, there is a finite number or web pages connected by links, as depicted in Figure \ref{f5}, which represents the XSD type for a web site. 

There is a mandatory XML attribute in the definition of a web site named base. This attribute is the starting URL for the navigation. The automation of the browsing will be carried out from this URL. Each page is recognised by a unique identifier. Each state can contain a set of data fields, and each data field contains the following information:

\begin{itemize}
\item Id: Data field identifier.
\item Locator: Optional identifier used to make a reference to a specific path (\textit{to} element within a \textit{transition}).
\item Type: Data type. It corresponds to the following HTML input elements: \textit{text}, \textit{textarea}, \textit{password}, \textit{checkbox}, \textit{radio}, \textit{file}, \textit{select-one}, and \textit{select-multiple}.
\item Required: Boolean value than indicates whether or not the data field is mandatory.
\item Value: Collection of values of the data field.
\item Stereotype: One of the following types: \textit{email}, \textit{date}, \textit{name}, \textit{surname}, \textit{address}, \textit{string}, \textit{integer}.
\end{itemize}

Moreover, each state can contain a set of oracles, which perform assertions in this web page. These oracles are described using the following attributes:

\begin{itemize}
\item Id: Oracle identifier.
\item Locator: Optional reference. It has the same meaning as in a data field.
\item Type: Oracle category. It can be one of the following literals: \textit{text} (assertation for a text to be present in the locator element), \textit{notText} (the opposite of \textit{text}), \textit{textPresent} (assertation for a text to be present in web page), \textit{textNotPresent} (the opposite of \textit{textPresent}), \textit{value} (assertation for a value to be present in the locator element), and \textit{notValue} (the opposite of \textit{value}).
\end{itemize}

Finally, web transitions are composed by an attribute called from (which is the identifier of the web page source) and a collection of actions and web targets (attribute \textit{to}). The action attributes is composed by the fields target, key, and event. The meaning of these fields is the same as in the guard of UML activity diagrams.

A simple example of navigation based on this XSD-schema is illustrated in the following snippet:

%\begin{lstlisting}[language=xml, style=eclipse,caption={XML Navigation Example}, label=l2]
\begin{lstlisting}[language=xml, style=eclipse]
<?xml version="1.0" encoding="ISO-8859-1" ?>
<website xmlns="http://www.dit.upm.es/atp" xmlns:xsi="http://www.w3.org/2001/XMLSchema-instance" xsi:schemaLocation="http://www.dit.upm.es/atp http://atestingp.sourceforge.net/atp.xsd" base="http://localhost:8080/WebAdmin/">
	<home id="login">
		<data locator="username">
			<value>Administrador</value>
		</data>
		<data locator="password">
			<value>admin</value>
		</data>
	</home>
	<transition from="login">
		<action target="frmDatos_0" event="click"/>
		<to state="init"/>
		<to state="login"/>
	</transition>
	<state id="init">
		<assert locator="texto-entrada" type="text">
			<value>Welcome</value>
		</assert>
	</state>
</website>
\end{lstlisting}

All in all, the approach we propose to automate the functional testing for web applications can be seen as an aggregation of the following automated methods:

\begin{itemize}
\item R\&P. Linear scripts using a record and playback method is used. This approach is considered the 1\textsuperscript{st} generation of AST frameworks.
\item Data-driven approach (2\textsuperscript{nd} generation). This testing approach means that using a single test case driving the test with input and expected values from an external data source instead of using the same hard-coded values each time the test runs.
\item MBT (3\textsuperscript{rd} generation). UML models from design phases (use cases, activity, and presentation diagram) will be reuse to guide the automation approach. 
\end{itemize}

\begin{figure}
  \caption{Schematic Diagram of the Test Case Automation}
  \label{f6}
  \begin{center}
  	\includegraphics[width=360px]{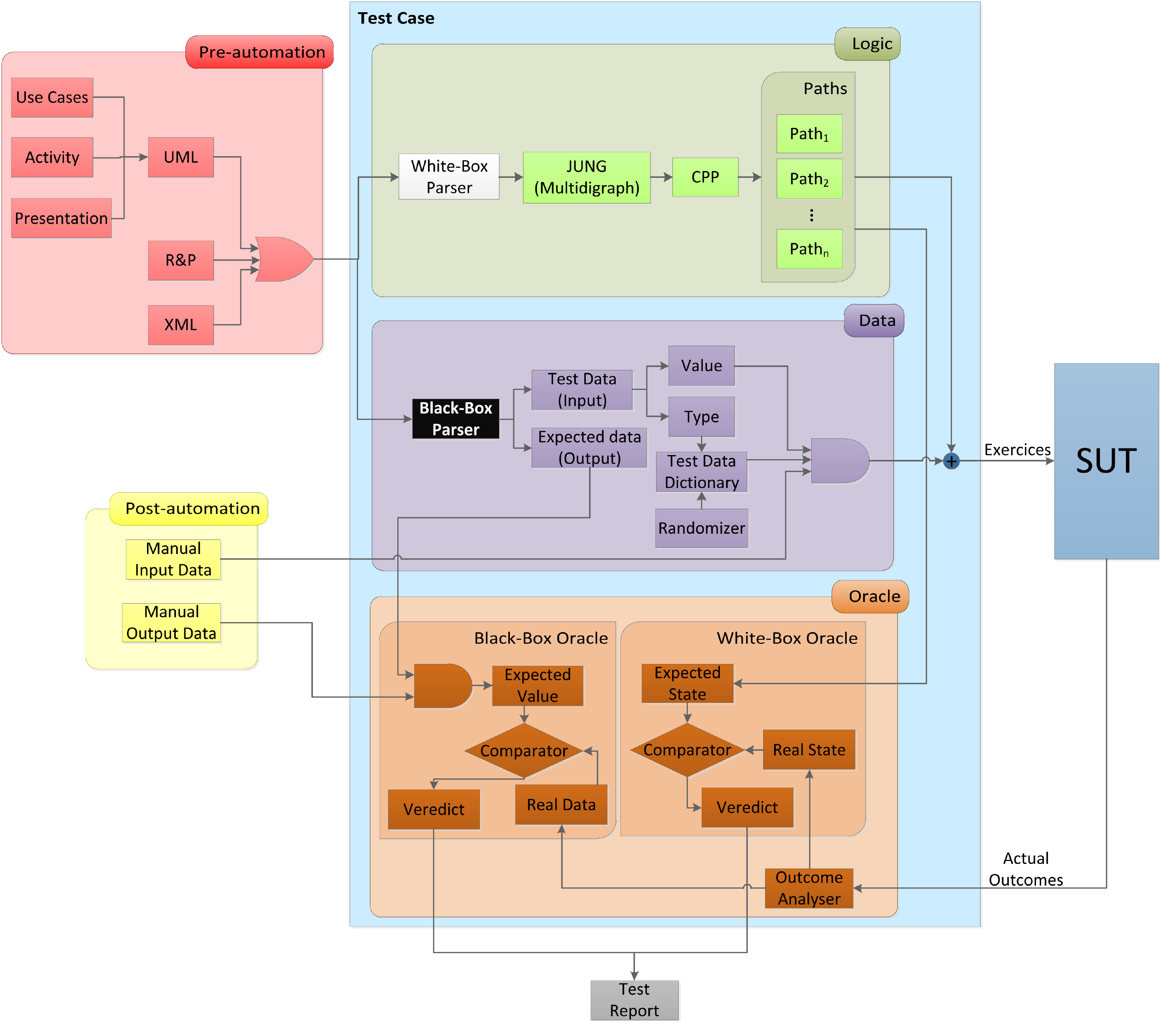}
  \end{center}
\end{figure}

In order to achieve the data-driven approach, the automation will mean the separation of the test case and test data/expected outcome generation. In order to store the test data and expected outcome a tabular data file will be used. This file will store test data (input) and expected outcomes (output). Therefore, this method has one strong prerequisite: there should be a model of the navigation behaviour of the web under test. As depicted before, this navigational model is one of these three notations:  UML (using NDT), or XML, or R\&P. This requisite is labelled as pre-automation in the red box illustrated in Figure \ref{f6}.

Once test cases for the navigation paths are generated, additional input and output data can be manually added to drive more test cases with the same test logic. These data (input and output) can be stored as new files in the tabular file as depicted before. This process is shown schematically in the yellow box labelled as post-automation in Figure \ref{f6}.

Regarding test case generation, the automation is done in three different stages: i) Test logic generation; ii) Test data generation; iii) Test oracle generation. Test logic generation is illustrated in the green box in Figure \ref{f6}. This step takes as input the model from the pre-automation stage, i.e. a model in UML, or XML or R\&P. This logic generation pass through the following steps:

\begin{itemize}
\item White-Box Parser. This entity is in charge of translating the different models used for testers (UML, XML, or R\&P) to the internal way of modelling web applications, that is, multidigraphs.
\item CPP. This module contains the deterministic CPP Java algorithm created by Thimbleby \cite{26}, used to find the paths within the multidigraph.
\item Paths. As a result of applying the CPP algorithm, a set of independent paths should be found. These paths correspond to a sequence of web pages that should be exercised against the SUT to ensure the navigation requirements.
\end{itemize}

Test data generation is illustrated in the purple box on Figure \ref{f6}. This stage is also fed with the navigation model from the pre-automation stage. The process is as follows:

\begin{itemize}
\item Black-Box Parser. This module extracts test data and expected outcome from the input model. XML models can include test data and R\&P models can include test data and expected outcome. Regarding test data (input), this black-box parser should extract the value and the data type. 
\item Data type will feed a test data dictionary. This dictionary contains a collection of data that can be used as input for test cases. For the selection of specific value, besides the type of data, a module that generates a random pointer will be used (randomizer).
\item Therefore, the data required for the test cases consist on the aggregation of three different sources: i) Data from the XML and R\&P models; ii) Randomly generated data from on a test data dictionary; iii) Manual data included as new rows in the tabulated file (post-automation).
\end{itemize}

Test oracle generation is illustrated in the green box on Figure \ref{f6}. This module has the following parts:

\begin{itemize}
\item Outcome analyser. This module collects data from the response of the SUT and extracts the following information: i) Navigation state; ii) Actual data returned by the application. In order to find out the real state navigation, the aggregation of data field will be used. 
\item White-Box Oracle. This module will establish verdicts by comparing the expected to the actual state. The expected state is set by the navigation path previously extracted in the test logic module. The real state is extracted from the SUT's response by the outcome analyser using the procedure described before (aggregation of data fields).
\item Black-Box Oracle. This module will establish verdicts by comparing expected with actual data. The expected data comes from the black-box parser of the test data module. In addition, additional expected data can be added in the post-automation stage by adding new information in the tabular data file.
\item Verdicts from white and black-box oracles will become the test report.
\end{itemize}

\section{Automatic Testing Platform}
\label{s5}

The tool which implements the testing proposed in this paper has been named Automatic Testing Platform (ATP)\footnote{\href{http://atestingp.sourceforge.net/}{http://atestingp.sourceforge.net/}} and has been released as open-source under the terms of Apache license 2.0. ATP has been built using existing open-source components, summarized in Table \ref{t2}. 

\begin{table}
  \caption{ATP Components}
  \label{t2}
  \begin{center}
    \begin{tabular}{ | l | l | l | } \hline 
\textbf{Function} & \textbf{Library} & \textbf{URL} \\ \hline
Unit framework & JUnit & \href{http://www.junit.org/}{http://www.junit.org/}\\ \hline
Web browsing & Selenium & \href{http://seleniumhq.org/}{http://seleniumhq.org/}\\ \hline
Test case generation & Freemarker & \href{http://freemarker.sourceforge.net/}{http://freemarker.sourceforge.net/}\\ \hline
Random data generation & dgMaster & \href{http://dgmaster.sourceforge.net/}{http://dgmaster.sourceforge.net/}\\ \hline
Test case execution & Ant & \href{http://ant.apache.org/}{http://ant.apache.org/}\\ \hline
Test case reporting & iText & \href{http://itextpdf.com/}{http://itextpdf.com/}\\ \hline
Graph manipulation & JUNG & \href{http://jung.sourceforge.net/}{http://jung.sourceforge.net/}\\ \hline
XML parsing & JDOM & \href{http://www.jdom.org/}{http://www.jdom.org/}\\ \hline
Spread-sheet access & JExcelAPI & \href{http://jexcelapi.sourceforge.net/}{http://jexcelapi.sourceforge.net/}\\ \hline 
    \end{tabular}
  \end{center}
\end{table}

\begin{figure}
  \caption{ATP Process}
  \label{f7}
  \begin{center}
  	\includegraphics[width=300px]{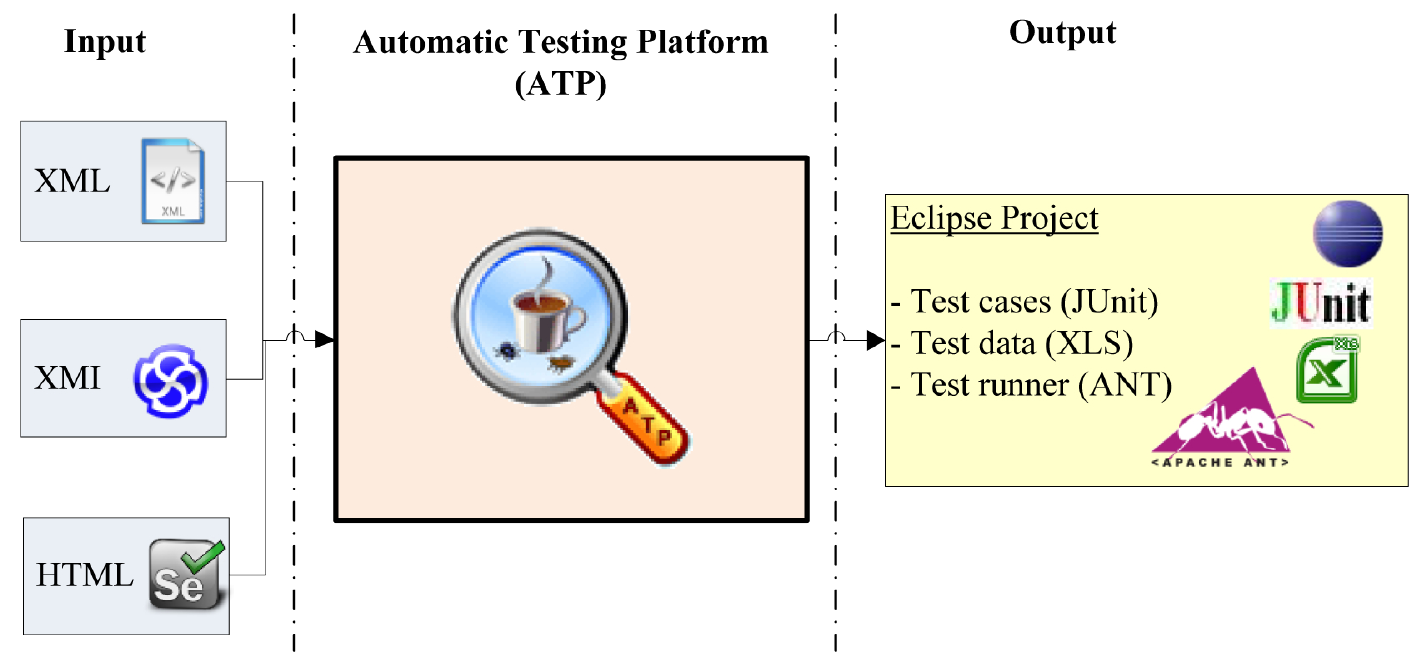}
  \end{center}
\end{figure}

Therefore, ATP accepts three kinds of inputs: i) XML navigation; ii) NDT files (UML); iii) Selenium scripts in HTML format. Regarding NDT approach, it uses Enterprise Architect (EA) to build its models \cite{19}. ATP accepts this EA models in XMI (XML Metadata Interchange) format. Regarding output, ATP creates a Java Eclipse project from the scratch with the following components inside (see Figure \ref{f7}):

\begin{itemize}
\item JUnit test cases. One per path.
\item Tabular test data (input and ouput). In an Excel spread-sheet per path.
\item Script runner (Apache Ant). This script starts the Selenium server before running the unit test cases.
\end{itemize}

ATP is a command-line tool. It has five main commands. Typing only ``atp'' in the shell, it shows the following help:

%\begin{lstlisting}[language=Ant, caption={ATP in the Shell}, label=l3]
\begin{lstlisting}[language=Ant]
> atp
[INFO] ATP (Automatic Testing Platform) v2.0
[INFO] [http://atestingp.sourceforge.net]
[INFO] Copyright (c) 2011 UPM. Apache 2.0 license.
[INFO]
[INFO] Use one of these options:
[INFO]          atp create
[INFO]          atp run
[INFO]          atp clean
[INFO]          atp list
[INFO]          atp set <key> <value>
[INFO]          atp report
\end{lstlisting}

The explanation of these commands is:
\begin{itemize}
\item Create: this command creates the test case and data for each path. The Eclipse project which contains these artifacts is also created with this command.
\item Run: this command executes the previously created test cases, by using the Ant script executor already created. Previous to the execution, a Selenium server is launched. As a result, test reports in different formats are created (XML, HTML, and PDF).
\item Clean: This command drops the Eclipse project previously created.
\item List: This command shows the configuration parameters of ATP. The most important parameters are: sut, which is the URL of the web under test; root, folder where the Eclipse project with all the output artifact to be created; navigation-type, type of input (xml, xmi or html); navigation-folder, root to the input file(s).
\item Set: This command change the value of any configuration parameter. For example: atp set navigation-type xmi.
\item Report: This command opens the HTML reports previously generated.
\end{itemize}

\section{Case Study: Management of Electronic Invoices}
\label{s6}

ATP's method and implementation has been validated using a complete web application developed for the Spanish company Telvent called ``Factura'' created in the context of IT Factur@ innovation project. This application is an electronic invoice web management system which has been developed using a Model Driven Engineering (MDE) approach \cite{29}. The Research Questions (RQ) which have driven this case study are the following: i) RQ1: Does the Factura application accomplish its functional requirements? ii) RQ2: Is ATP capable of finding defects in a finished web application? iii) RQ3: What are the advantages and disadvantages of different types of input (UML, XML, and R\&P) to ATP? 

The input model for ATP will be the XMI models created using EA following the NDT approach. 5 use cases have been identified, and each use case has been refined using an activity diagram. ATP uses the information of each activity diagram as input to create an equivalent navigation graph. Figure \ref{f8} shows the use case diagram and an example of activity diagram:

\begin{figure}
  \caption{Use Cases and Activity Diagrams}
  \label{f8}
  \begin{center}
  	\includegraphics[width=450px]{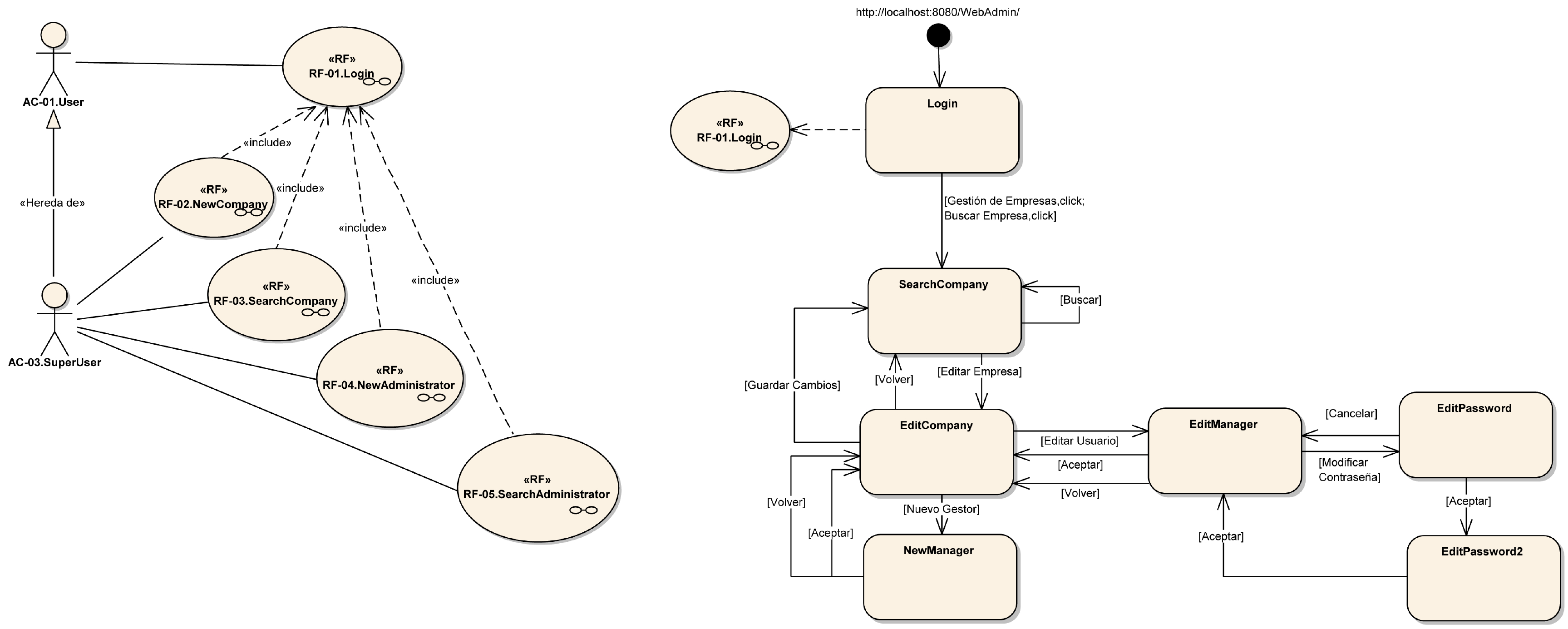}
  \end{center}
\end{figure}

\begin{figure}
  \caption{Case Study Report}
  \label{f9}
  \begin{center}
  	\includegraphics[width=450px]{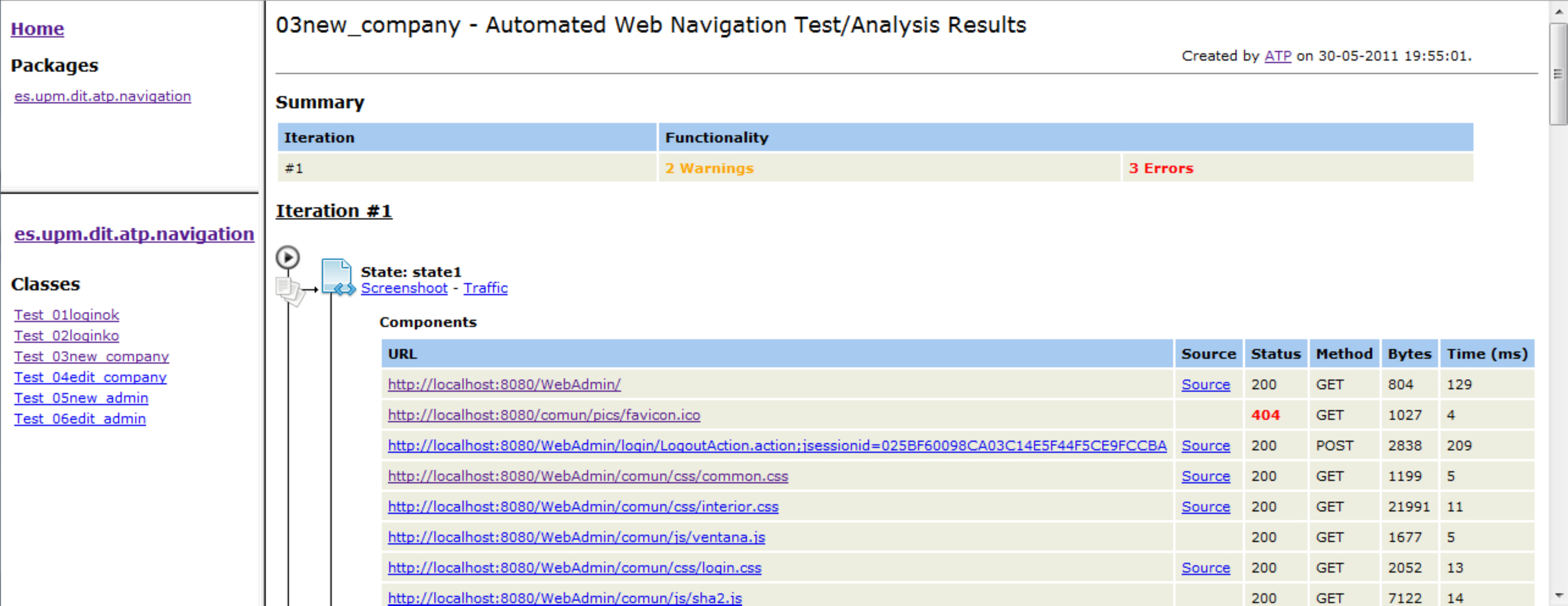}
  \end{center}
\end{figure}

After that and using the CPP algorithm, it breaks the graph into its path, which is written as unit test cases (JUnit). Test data is also created using the information attached to the model when possible, and using random data from test data dictionaries otherwise. Finally, the test cases were executed. As a result, 6 navigation errors have been found. In addition, 8 JavaScript notifications were reported by ATP. These notifications include JavaScript alerts, prompts, and confirmations. An snapshot of the generated report is illustrated in Figure \ref{f9}.

Regarding RQ1, we can conclude that Factura is quite good implemented since the navigation error number is low. Regarding RQ2, it has been proved that ATP can be a useful tool to discover functional failures in web applications. RQ3 is about the way that ATP works. The pros and cons of these inputs are summarized in Table \ref{t3}.

\begin{table}
  \caption{ATP Pros and Cons}
  \label{t3}
  \begin{center}
    \begin{tabular}{ | p{2cm} | p{6cm} | p{6cm} | } \hline 
\textbf{Input} & \textbf{Pros}  & \textbf{Cons} \\ \hline
XMI (UML models in NDT) &  Analysis/design models are reused for assessment. Every possible path is depicted in the models. & It is not possible to attach test data nor oracle in the models. Post-automation step is mandatory.   \\ \hline
XML (based on XSD Schema) &  Every possible path can be depicted using XML files. Data and oracles can be attached to XML files & The XML files must be coded and maintained by hand.  \\ \hline
HTML (Selenium R\&P scripts) &  The creation of the scripts is done using Selenium IDE against the real application. Data and oracles can be attached to HTML scripts & Each recording is linear, therefore there isalways a single path by HTML script.	Error paths should be defined in different scripts. \\ \hline
    \end{tabular}
  \end{center}
\end{table}

\section{Conclusion}
\label{s7}

This paper has presented a method for the automated testing of web applications based on their navigation. The basic idea behind this approach is to exercise the SUT using a real browser, performing the navigation from page to page by means of web links. During this navigation, a functional validation is carried out since the correctness of the links and the underlying logic is executed. Each web page corresponds to a state in the navigation picture, which is ensured to be correct during the navigation. 

The first kind of input for this method can be UML models of the SUT. Concretely, three kinds of UML diagrams are needed: use case, activity, and presentation diagram. Since presentation diagrams are not standard in UML 2.0, we have done a study of the state-of-the-art in web modeling, and we have decide to use NDT as our choice to model web application using UML. The second alternative in our method is using a R\&P script recorded using Selenium IDE, since Selenium RC will be the tool in charge of the automation of the web navigation. Third and last, an XML-based file for the navigation can be employed. This XML file is a simple way of structuring the navigation following a XSD schema.

This approach has been implemented in an open-source framework named Automatic Testing Platform (ATP). This tool automates the testing in four levels: i) test case logic generation; ii) test data derivation; iii) test oracle generation; iii) test case execution driven by Ant scripts; iv) test case reporting. To validate the proposed method, a web application for invoice management system has been employed. ATP has proven to be capable to find functional defects in a real web application.

Future work will extend the presented approach through the automation of testing and analysis of non-functional requirements such as performance, security, compatibility, usability and accessibility.

\section*{Acknowledgment}

This paper has been performed in the context of the European project ITEA-MOSIS (project number 06035), under grant by Spanish Ministerio de Industria, Turismo y Comercio in the PROFIT program.

\bibliographystyle{eptcs}

\end{document}